\documentclass{ws-procs9x6}
\usepackage{epsfig}
\input epsf.sty

\newcommand{\nwc}{\newcommand}

\nwc{\be}  {\begin{displaymath}}
\nwc{\ee}  {\end{displaymath}}
\nwc{\bmu} {\bar{\mu}}
\nwc{\ba}  {\begin{eqnarray*}}
\nwc{\ea}  {\end{eqnarray*}}
\nwc{\nn}  {\nonumber\\}
\nwc{\tr}  {\mathop{\rm Tr}}
\nwc{\Tr}  {\mathop{\rm Tr}}
\nwc{\re}  {\mathop{\rm Re}}
\nwc{\im}  {\mathop{\rm Im}}
\nwc{\Hc}  {\mathop{\rm H.c.}}
\nwc{\la}[1]{\label{#1}}
\nwc{\rmi}[1]{{\mbox{\scriptsize #1}}}
\nwc{\nr}[1]{(\ref{#1})}
\nwc{\fr}[2]{{\frac{#1}{#2}}}
\nwc{\msbar}{\overline{\mbox{\rm MS}}}
\nwc{\lambdamsbar}{\Lambda_{\overline{\rm MS}}}
\newcommand{\fig}{Fig.~}
\newcommand{\eq}{Eq.~}

\newcommand{\eqs}{Eqs.~}

\def\lsi{\raise0.3ex\hbox{$<$\kern-0.75em\raise-1.1ex\hbox{$\sim$}}}
\def\gsi{\raise0.3ex\hbox{$>$\kern-0.75em\raise-1.1ex\hbox{$\sim$}}}
\nwc{\lsim}{\mathop{\lsi}}
\nwc{\gsim}{\mathop{\gsi}}

\begin{document}


\title{What is the Simplest Effective Approach\\ to Hot QCD Thermodynamics?}

\author{M. LAINE}

\address{Theory Division, CERN, CH-1211 Geneva 23, Switzerland}

\maketitle

\abstracts{The dimensionally reduced action is believed to provide 
for a theoretically consistent and numerically precise effective 
description of the thermodynamics of the quark-gluon plasma, once 
the temperature is above a few hundred MeV. Although dramatically 
simpler than the original QCD it is, however, still a strongly 
interacting, confining theory. In this talk 
I speculate on whether there could exist a further simplified 
recipe within that theory, for physically relevant temperatures, 
which would already lead to a phenomenologically satisfactory 
description of the free energy and various correlation lengths of hot QCD, 
but with only a minimal amount of numerical non-perturbative input needed.}

\section{Introduction}

It is hoped that one day the reliability of the computational methods 
developed for finite temperature QCD can be systematically checked against
experimental data from heavy ion collision experiments. Indeed, 
we could then apply the same tools to important problems 
in cosmology with some more confidence. 
A well-known example of an observable on which a lot remains 
to be understood under both circumstances, is the time scale at which 
a non-Abelian system thermalizes, starting from some complicated initial 
non-equilibrium state. 

We should therefore try to develop  methods
for determining physical observables at high temperatures which are
both theoretically consistent, numerically precise, and possibly also 
simple to apply. 
In this talk, this goal is
further narrowed in two ways. First of all, we only discuss 
temperatures $T$ sufficiently 
above the deconfinement transition (or crossover), 
$T \gsim 2 T_c^\rmi{QCD}$. Second, we only consider ``static'' 
observables, such as correlation lengths and the equation of state. 
In other respects we do in principle allow for the full physical QCD, with 
$N_f > 0$ light dynamical quark flavours and a small baryon chemical 
potential $\mu$, although for brevity the data shown 
below are restricted to $N_f = 0$, $\mu/T=0$.

Since our ultimate interest 
is not only to be theoretically consistent but also
numerically precise, we choose for the sake of this talk a rather 
phenomenological approach. 
That is, for lack of an {\it a priori} systematic analytic framework
for describing confining dynamics, we will compare the results of 
a number of recipes to be introduced presently with actual data, although
not from collider but from lattice experiments. Thus we    
obtain evidence for some conjectures concerning the non-perturbative 
dynamics. 

There are also other approaches on the market with 
somewhat similar objectives; for recent reviews see, 
e.g., Refs.~\cite{other,revs,ma}. 
Their general framework is however
very different from ours, with the exception perhaps of Ref.~\cite{ck}, 
which also takes the dimensionally reduced theory as a starting point. 

\section{Physical scales of the system}

\subsection{The asymptotic regime}
\la{se:asymptotic} 

Let us start by recalling the parametric scales appearing in finite
temperature QCD in the static limit. The periodicity over the Euclidean
time direction introduces the scale present already in a non-interacting 
theory, $p\sim 2 \pi T$. For the bosonic Matsubara zero modes which are 
massless in the non-interacting theory, interactions introduce 
softer scales, corresponding to collective plasma modes. 
In particular, colour-electric fields
are screened at the Debye scale $p\sim gT$, while colour-magnetic 
fields are screened only non-perturbatively, at the scale 
$p\sim g^2 T$~\cite{linde,gpy}. Due to three-dimensional confinement, 
there are no scales softer than this, so that the list is exhaustive. 

As is well-known, perturbation theory has a number of problems
in this setting. This can perhaps most easily be understood from 
the fact that there is an expansion parameter related to bosonic 
fluctuations with momentum scale $p$ of the form
\begin{equation}
 \epsilon \sim 
 \frac{1}{\pi} g^2(2\pi T) n_b(p) = 
 \frac{g^2(2\pi T)}{\pi (e^{p/T}-1)} 
 \stackrel{p < T}{\sim}\frac{g^2(2 \pi T) T}{\pi p}, 
\end{equation}
where $n_b$ is the Bose-Einstein distribution function. Thus, for 
hard modes, $p \sim 2\pi T$, the expansion parameter is $\sim g^2/\pi^2$ 
like at $T=0$, and should remain viable down to a scale 
$2\pi T \sim 2$ GeV, which corresponds roughly to $T \sim 2 T_c^\rmi{QCD}$. 
For the colour-electric modes, $p\sim g T$, the series is  
only in $\sim g/\pi$, and seems to converge much worse, 
unless $T \gg T_c^\rmi{QCD}$~\cite{az,zk,bn,adjoint}. 
For colour-magnetic modes, 
$p\sim g^2 T$, there is no perturbative series~\cite{linde,gpy}.

Thus, there is certainly one simplification from the full theory one 
can make: at $T\gsim 2 T_c^\rmi{QCD}$, 
the scales $p\sim 2 \pi T$ can be integrated 
out. This leads to a dimensionally reduced effective theory
for the bosonic zero modes~\cite{dr}: 
\begin{equation}
 {\mathcal{L}}_\rmi{E} = 
 \fr12 \Tr F_{ij}^2
 + \Tr [D_i,A_0]^2 + 
 m_\rmi{E}^2\Tr A_0^2 +\lambda_\rmi{E}(\Tr A_0^2)^2. 
 \label{SE}
\end{equation}
Here $m_\rmi{E}^2 \sim g^2T^2$, $\lambda_\rmi{E} \sim g^4 T$ are generated 
radiatively~\cite{dr}. These parameters, as well as the effective
gauge coupling, have been computed up to next-to-leading order
(Ref.~\cite{adjoint} and references therein). The question
we address in this talk is whether any further simplifications
are possible. 

In the limit of asymptotically high temperatures,
corresponding to $g\ll 1$, 
the different scales in the system come with the hierarchy
\begin{equation}
 g^2 T \ll gT \ll 2 \pi T. \la{hier_asy}
\end{equation}
Then the action in~\eq\nr{SE} can certainly be simplified, by 
integrating out~$A_0$. One consequence is that all
local gauge-invariant operators which do not have quantum numbers 
forbidding this, have correlation lengths of the
order of the magnetic scale, $\mathcal{O}((g^2 T)^{-1})$~\cite{ay}.  

\subsection{The real world}
\la{se:real}

For realistic temperatures, however, the situation is different. 
In fact, the longest correlation lengths are associated with operators
such as $\tr A_0^2$, and the corresponding mass scale is significantly
below that for purely magnetic objects~\cite{hp}
(see also below). Hence the 
realistic ordering is, in some sense,  
\begin{equation}
 gT \ll g^2 T \ll 2 \pi T. \la{hier_real}
\end{equation}
It is easy to understand that this has to be the case:
close to $T_c^\rmi{QCD}$ the dynamics starts to be dominated by the Z($N_c$)
symmetry related to the Polyakov loops,
denoted here by $\mbox{Pol}$ (for recent 
literature, see Ref.~\cite{rdp} and references therein), 
which are represented in the effective theory 
just by $\tr A_0^2 \sim \re\tr\mbox{Pol}, 
\tr A_0^3 \sim \im \tr \mbox{Pol}$. 
(Note that the discussion here is in terms of degrees
of freedom represented by spatially local gauge-invariant operators, 
rather than constituents~\cite{op}, although this does
not really make much of a difference). 

We may thus state, at least 
on some heuristic level, that:
\begin{itemize}
\item
 Since the magnetic scale $\sim g^2T$ is large
 (cf.\ \eq\nr{hier_real} and \fig\ref{fig:comparison} below), 
 it has to be taken into
 account in all observables where it enters additively.
\item
 At the same time, correlation lengths which are parametrically 
 perturbative ($\sim gT$),
 remain so non-perturbatively too, 
 because the magnetic glueball is too heavy to
 enter as an intermediate state.
\end{itemize}

\section{Conjecture}
\la{se:conj}

Motivated by this physical picture, we propose as an organizing 
principle for the remainder of this presentation the following conjecture: 
\begin{itemize}
\item[]
If one starts from the dimensionally reduced theory 
and determines physical observables 
up to and including the leading non-perturbative ``magnetic'' 
contribution, such that effects from all the scales have made their 
entrances, then one should already have accounted for most of 
the dynamics at the scales $\sim g T,g^2T$.
\end{itemize}
In the remainder, we discuss numerical evidence in favour
of this proposal.

\section{Observables}

The observables to be discussed, 
and their parametric behaviours, are: 
\begin{itemize}
\item
The spatial string tension $\sqrt{\sigma_s}$:
\begin{equation}
 \sqrt{\sigma_s} \sim \mbox{[non-pert]} \times g^2 T + ...\;.
 \la{ss}
\end{equation}
\item
Correlation lengths $\xi_\rmi{G} \equiv m_\rmi{G}^{-1}$ 
related to ``eigenstates'' $\sim \Tr F_{ij}^2$:  
\begin{equation}
 m_\rmi{G} \sim \mbox{[non-pert]} \times g^2 T + ...\;.
 \la{mG}
\end{equation}
\item
Correlation lengths $\xi_\rmi{E}^{(i)} \equiv {[m_\rmi{E}^{-1}]}^{(i)}$
for $\re \Tr\, $Pol, $\im \Tr\,$Pol, $\Tr F_{0i} F_{jk}$,
etc: 
\begin{equation}
 [m_\rmi{E}]^{(i)} 
 \sim g T + g^2 T (\ln \frac{1}{g} + \mbox{[non-pert]}) + ...\;. 
 \la{mE}
\end{equation}
\item
Pressure, energy density, ...$\;$: 
\begin{equation}
 \frac{p}{T^4} \sim 
 1 + g^2 + g^3 + g^4\ln\frac{1}{g} + 
 g^5 + g^6 (\ln\frac{1}{g} + \mbox{[non-pert]} ) + ...\;. 
 \la{p}
\end{equation}
\end{itemize}
The conjecture then says that if we determine
all the terms shown in \eqs\nr{ss}--\nr{p}, 
then the sum of the 
remaining corrections is maybe on the $\mathcal{O}(20\%)$ 
level or so, rather than of order unity.

Let us try to clarify one aspect of the procedure. 
Indeed, the recipe introduced is of course the same as could be
used for asymptotically small values of $g$, as in 
Sec.~\ref{se:asymptotic}. However, in that case the 
non-perturbative correction would be small. For the 
``real world'' case of Sec.~\ref{se:real}, on the other hand, 
this procedure is just a ``recipe'': the expansion is formally
not sound, since $g$ is not that small and 
the correction is huge, and it is only our conjecture which
asserts that the {\it sum} of all further terms is subdominant. 

How are the series shown in \eqs\nr{ss}--\nr{p} to be determined?
This depends on the observable. Formally, one has to systematically 
integrate out $A_0$ to a sufficient depth, and measure the remaining 
non-perturbative numerical factor, using the theory 
\begin{equation}
 {\mathcal{L}}_\rmi{M} = \fr12 \Tr F_{ij}^2 \;.
 \la{SM}
\end{equation}
This is very simple for $\sqrt{\sigma_s}$ and $m_\rmi{G}$, which are 
obtained from finite observables within the theory in~\eq\nr{SM}, so 
that one only needs to compute the change in the effective coupling 
constant $g$.  

For the other observables one in general needs an explicit 
matching computation, because the operators to be used within 
\eq\nr{SM} are ultraviolet divergent. For instance, for electric
screening lengths the structure is~\cite{ay}
\begin{eqnarray}
[m_\rmi{E}]^{(i)} & \equiv &  - \lim_{z\to\infty} \frac{1}{z} \ln
\langle
{\mathcal{O}}_\rmi{E}^{(i)}(z)
{\mathcal{O}}_\rmi{E}^{(i)}(0)
\rangle_{{\mathcal{L}}_\rmi{E}} \\ 
 & = &  
 gT + g^2T \ln \frac{gT}{\Lambda_\rmi{M}} 
 - \lim_{z\to\infty} \frac{1}{z} \ln
\langle
{\mathcal{Q}}_\rmi{M}^{(i)}(z;0)
\rangle_{{\mathcal{L}}_\rmi{M},\Lambda_\rmi{M}} 
+ {\mathcal{O}}(g^3T) , \hspace*{0.3cm} 
\la{mE2} 
\end{eqnarray}
where $\Lambda_\rmi{M}$ denotes some ultraviolet regulator within 
the theory in~\eq\nr{SM}, and ${\mathcal{Q}}_\rmi{M}^{(i)}(z;0)$
is a specific non-local operator, which depends on the corresponding
operator ${\mathcal{O}}_\rmi{E}^{(i)}(z)$
in the theory of~\eq\nr{SE}. The ``non-pert'' term in
\eq\nr{mE} arises as a combination of the 2nd and 3rd terms
on the right-hand-side of~\eq\nr{mE2}. 

Finally, one can carry out a similar matching computation for $p(T)$. 
The first step in the procedure, dimensional reduction, leads to 
\begin{eqnarray}
 \frac{p}{T^4} & = &  1 + g^2 + g^4\ln\frac{2\pi T}{\Lambda_\rmi{E}} 
  + g^6\ln\frac{2\pi T}{\Lambda_\rmi{E}} + ... + \frac{p_\rmi{E}}{T^4},
 \la{LamE} \\ 
 \frac{p_\rmi{E}}{T^4} & = & \lim_{V\to\infty} \frac{1}{V T^3} 
 \ln \int \! {\mathcal{D}}A_i {\mathcal{D}}A_0  
 e^{-\int d^3x {\mathcal{L}}_\rmi{E}}.
 \la{p1} 
\end{eqnarray}
Here $V$ is the volume, and $\Lambda_\rmi{E}$ is 
the ultraviolet cutoff within the theory of~\eq\nr{SE}. 
The second part, integration over $A_0$, leads to
\begin{eqnarray}
\frac{p_\rmi{E}}{T^4} & = &  
 g^3 + g^4\ln\frac{\Lambda_\rmi{E}}{m_\rmi{E}} 
 + g^5 + g^6\ln\frac{\Lambda_\rmi{E}}{m_\rmi{E}}
 + g^6\ln\frac{m_\rmi{E}}{\Lambda_\rmi{M}} + ... + \frac{p_\rmi{M}}{T^4},
 \la{LamM} \\  
 \frac{p_\rmi{M}}{T^4} & = & \lim_{V\to\infty} \frac{1}{V T^3} 
 \ln \int \! {\mathcal{D}}A_i 
 e^{-\int d^3x {\mathcal{L}}_\rmi{M}}.
 \la{p2}
\end{eqnarray}
The terms here were
determined up to order ${\mathcal{O}}(g^5)$ in Ref.~\cite{bn}, 
and the coefficients of the ${\mathcal{O}}(g^6)$ logarithmic
terms are determined in~\cite{sd,gsixg,prep}. 

Unfortunately, unlike in~\eq\nr{mE2}, the numerical factors inside
the logarithms in \eqs\nr{LamE}, \nr{LamM} have not been computed yet. 
This means that the ``non-pert'' term of~\eq\nr{p} cannot be 
resolved. Since such numerical factors correspond however just 
to a constant, we can treat this constant as a free parameter, 
and see whether the conjecture has any chance of surviving. 
Ultimately, the constant is computable by a collection
of various techniques, as outlined in Ref.~\cite{gsixg}. 

\section{Evidence}

We now turn to data. 
When we discuss the observables in \eqs\nr{ss}--\nr{p}, 
we will in general refer to four different kinds of approximations:
\begin{itemize}
\item[(1)]
``Full 4d'':
Results from full 4d lattice Monte Carlo simulations. 

\item[(2)]
``Full 3d'':
Results from 3d lattice Monte Carlo simulations of the action
in~\eq\nr{SE}, with the effective parameters expressed in terms of 
the physical 4d parameters and $T$ as specified 
in Ref.~\cite{adjoint}.

\item[(3)]
``Leading perturbative term'':
For expressions of the form in~\eq\nr{mE}, the leading contribution
of order $gT$, given by some multiple of the parameter $m_\rmi{E}$
of the action in~\eq\nr{SE}. 

\item[(4)]
``Up to first non-perturbative term'': 
Terms up to the order shown in~\eqs\nr{ss}--\nr{p}, in cases where 
this part has been resolved from the ``full 3d'' result, for instance
through a procedure such as the one shown in~\eq\nr{mE2}.
\end{itemize}
As is well-known, the ``full 3d'' and ``full 4d'' results
for correlation lengths agree 
almost within statistical errors, and in any case after estimating 
the magnitude of systematic errors
arising from lattice artifacts~\cite{hp}. Our conjecture
then claims that the 4th approximation, which requires only a single
number of non-perturbative input 
(determinable with the theory of~\eq\nr{SM}) 
rather than full 3d
simulations of the more complicated action in~\eq\nr{SE}, 
should produce an outcome close to the values of the first two 
procedures. 

\subsection{Spatial string tension and correlation lengths}

The numerical data available for the spatial string tension
$\sqrt{\sigma_s}$; for the inverse of the correlation length
related to the eigenstate which connects 
to that determined from $\tr F_{ij}^2$ at asymptotically 
high $T$, denoted here by $m_\rmi{G}$; and for
inverses of correlation lengths
which have a perturbative leading term from the scale $gT$, 
denoted by $[m_\rmi{E}]^{(i)}$; 
are shown in~\fig\ref{fig:comparison}.

We observe that
for purely magnetic observables ($\sqrt{\sigma_s}$, $m_\rmi{G}$), 
results obtained with~\eq\nr{SM} agree nicely with full 3d 
results, as well as with full 4d results, where available. In particular, 
even though in principle there, there is no need to include any
higher order corrections beyond those determined by~\eq\nr{SM}. 
This observation is certainly in perfect agreement with our 
conjecture.

For the electric observables, on the other hand, 
the ``up to first non-pert.\ 
term''-prediction has so far been resolved in only one quantum 
number channel, although it could in principle be resolved in others, 
as well. We no longer observe perfect agreement. However, 
while the leading perturbative terms are too small by a factor $\sim 3$, 
``up to first non-pert.\ term'' 
predictions are too large by $\sim 30$ \%. Thus the 
conjecture still works 
semi-quantita\-tively: all the higher order corrections
in~\eq\nr{mE2} sum up to a small number, even though the first correction
beyond the leading term is huge. It would be interesting
to see whether the conclusion holds also after studying other
channels, where the non-perturbative correction should be much smaller, 
in order to fit the full 4d and full 3d results
(cf.\ \fig\ref{fig:comparison}).

\begin{figure}[t]


\centerline{\epsfxsize=10cm\epsfbox{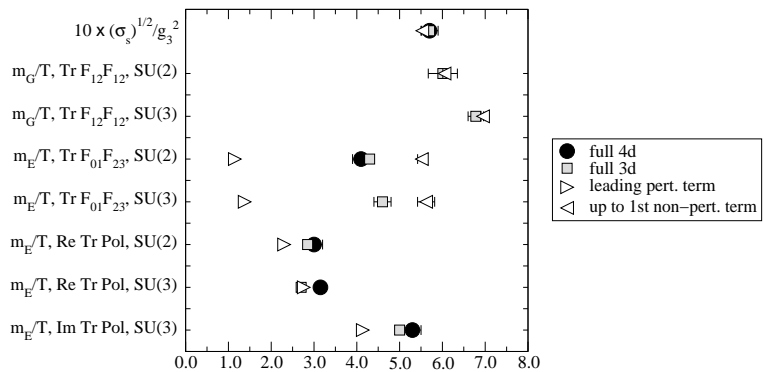}}

\vspace*{0.5cm}

\caption[a]{A comparison of spatial string tensions, as well as 
inverses of correlation lengths in various channels
and two gauge groups, at $T\approx 2 T_c^\rmi{QCD}$, with different methods, 
as explained in the text. The lattice data have been collected from 
Refs.~\cite{hp,bi,mt,lp,dg}. Representative continuum operators
have also been shown.}

\la{fig:comparison}
\end{figure}

\subsection{Pressure}

We finally turn to the pressure. Unfortunately the situation 
remains somewhat open here. Not only has the non-perturbative $g^6$ 
term in~\eq\nr{p} not been resolved, but we also do not have 
a ``full 3d'' prediction available: measurements can be 
carried out but a constant again remains open~\cite{a0cond,latt}. 
These problems are ultimately
due to the fact that the constant parts related to the divergent
4-loop matching coefficients in~\eqs\nr{LamE}, \nr{LamM} have not 
been determined yet. The only unambiguous results are from 4d lattice
simulations (at least for $N_f = 0$), and from perturbation theory
at asymptotically high temperatures. 

Therefore, the best we can do is to see whether a {\it necessary} 
condition for the validity of the conjecture is satisfied: is it possible
to find some value representing the unknown 4-loop constant terms, 
such that it reproduces {\it a whole function} of the expected type?
This exercise is carried out in~\fig\ref{fig:cdep}
(taken from Ref.~\cite{gsixg}), and we indeed
find that one can reproduce, in principle, a reasonable
curve with a value of the constant of order unity.

\begin{figure}[t]


\centerline{\epsfxsize=6cm\epsfbox{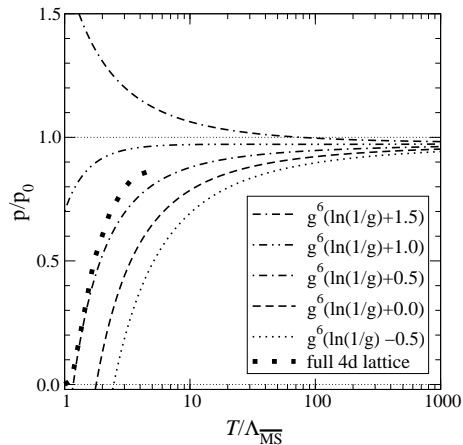}}


\caption[a]{The pressure, normalised to the non-interacting
Stefan-Boltzmann law $p_0$, for various values of the 
$\mathcal{O}(g^6)$ constant term (from Ref.~\cite{gsixg}; 
see there also for further details on what exactly is plotted here).
The 4d lattice data is from Refs.~\cite{bi,others}.}

\la{fig:cdep}
\end{figure}

Of course, it must be said that due to the high perturbative order
appearing, the pressure is an observable which is quite sensitive
not only to the non-perturbative term but also to ambiguities in 
perturbation theory. For illustration, we demonstrate this in
\fig\ref{fig:mudep}. An optimal value has been chosen for the 
constant, and then the renormalisation scale entering the 
expressions of the effective parameters in~\eq\nr{SE} has been varied 
within the range $0.5\bmu_\rmi{opt}...2.0\bmu_\rmi{opt}$, 
where $\bmu_\rmi{opt} \approx 6.7 T$ is suggested by the 
next-to-leading order expression for $g_\rmi{E}^2$~\cite{adjoint}.

The scale dependence is non-monotonic, because the first derivative
of the result with respect to $\bmu$ is very close to vanishing, 
due to the fact that we have
gone via the effective theory in~\eq\nr{SE} to obtain the output. 
Nevertheless, we see that at $T\lsim 3 T_c^\rmi{QCD}$ 
($T_c^\rmi{QCD}\sim \lambdamsbar$) the predictions are 
in any case getting unreliable, irrespective of the value of 
the constant. This may not be too surprising, though, since to determine
the full $g^6$ constant term requires also renormalisation of $g$
at 2-loop level, but this has not been incorporated in 
the expressions shown here. 

\begin{figure}[t]


\centerline{\epsfxsize=6cm\epsfbox{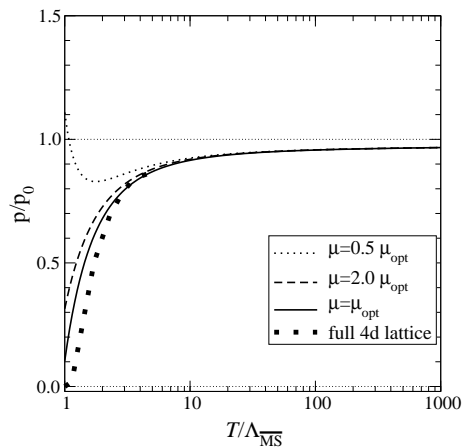}}

\vspace*{0.5cm}

\caption[a]{The dependence of the result of~\fig\ref{fig:cdep} on 
the renormalisation scale $\bmu$, around $\bmu_\rmi{opt}\approx 6.7T$, 
for an $\mathcal{O}(g^6)$ constant $\sim 0.7$ (cf.\ Ref.~\cite{gsixg}).}

\la{fig:mudep}
\end{figure}

\section{Conclusions}

There is some numerical evidence for the conjecture
of Sec.~\ref{se:conj} from the spatial string tension 
as well as from various (``magnetic'' and ``electric'')
correlation lengths, but more channels should be added. 
It would also be interesting to consider other observables 
(see, e.g., Ref.~\cite{gka}).

The conjecture might also work for the pressure, although
so far only one necessary condition is observed to be satisfied,
and a lot of work remains to be carried out before we know
whether a sufficient condition is there.
 
Once such analytic expressions with non-perturbative constants
are known, extensions to finite $N_f>0$ and $\mu\lsim \pi T$ are 
easy (see, e.g.,~Ref.~\cite{av}).  

Hopefully similar expressions can be worked out 
also for real time observables,
where 4d lattice simulations are even much more difficult. 

\section*{Acknowledgments}

I thank A.~Hart, K.~Kajantie, O.~Philipsen, K.~Rummukainen, 
Y.~Schr\"oder and M.~Shaposhnikov, for discussions and
collaboration on topics mentioned in this talk. 


\end{document}